\documentclass[prl,twocolumn,superscriptaddress,nofootinbib,longbibliography,tightenlines]{revtex4-2}
\usepackage[svgnames]{xcolor}
\usepackage{amsmath,amssymb,amsthm,bbm}
\usepackage{xr-hyper}
\usepackage{hyperref}
\usepackage{graphicx}
\usepackage{float}
\usepackage[capitalise]{cleveref}
\hypersetup{
  colorlinks,
  citecolor=Salmon,
  linkcolor=Teal,
  urlcolor=Teal}

\newcommand\red[1]{#1}
\newcommand\redeq[1]{#1}
\newcommand\blue[1]{}
\usepackage[normalem]{ulem}

\usepackage{slashed}
\usepackage{tikz}
\usepackage[compat=1.1.0]{tikz-feynman}
\usepackage{subcaption}
\usepackage{ragged2e}
\DeclareCaptionJustification{justified}{\justifying}
\newcommand{\SU}{\operatorname{SU}}
\renewcommand{\O}{\operatorname{O}}

\newcommand{\e}{\operatorname{e}}
\renewcommand{\d}{\mathrm{d}}
\renewcommand{\i}{\mathrm{i}}
\newcommand{\tr}{\operatorname{tr}}
\newcommand{\Tr}{\operatorname{Tr}}
\newcommand{\Det}{\operatorname{Det}}

\newcommand{\sgn}{\operatorname{sgn}}
\newcommand{\<}{\langle}
\renewcommand{\>}{\rangle}
\newcommand{\eff}{\textnormal{eff}}
\newcommand{\crit}{\textnormal{crit}}

\usepackage{makecell}

\makeatletter
\newcommand*{\addFileDependency}[1]{
  \typeout{(#1)}
  \@addtofilelist{#1}
  \IfFileExists{#1}{}{\typeout{No file #1.}}
}
\makeatother

\newcommand*{\myexternaldocument}[1]{%
    \externaldocument{#1}%
    \addFileDependency{#1.tex}%
    \addFileDependency{#1.aux}%
}
\myexternaldocument{SM}

\begin{document}

\title{Quantum Criticality of Anti-ferromagnetism and Superconductivity with Relativity}
\author{Hanqing Liu}
\affiliation{Department of Physics, Box 90305, Duke University, Durham, NC 27708, USA}
\author{Emilie Huffman}
\affiliation{Perimeter Institute for Theoretical Physics, Waterloo, ON N2L 2Y5, Canada}
\author{Shailesh Chandrasekharan}
\affiliation{Department of Physics, Box 90305, Duke University, Durham, NC 27708, USA}
\author{Ribhu K. Kaul}
\affiliation{Department of Physics \& Astronomy, University of Kentucky, Lexington, KY 40506, USA}
\begin{abstract}
We study a quantum phase transition from a massless to massive Dirac fermion phase in a new two-dimensional bipartite lattice model of electrons that is amenable to sign-free  quantum Monte Carlo simulations. Importantly, interactions in our model are not only invariant under $\SU(2)$ symmetries of spin and charge like the Hubbard model, but they also preserve an Ising like electron spin-charge flip symmetry. From unbiased fermion bag Monte Carlo simulations with up to 2304 sites, we show that the massive fermion phase spontaneously breaks this Ising symmetry, picking either anti-ferromagnetism or superconductivity and that the transition at which both orders are simultaneously quantum critical, belongs to a new ``chiral spin-charge symmetric" universality class. We explain our observations using effective potential and renormalization group calculations within the framework of a continuum field theory.
\end{abstract}

\maketitle

The study of graphene has triggered an avalanche of interest in the physics of 
massless relativistic fermions in two spatial dimensions, highlighting  the connections between condensed matter and high energy physics theory, and the unity of physics across disparate energy scales~\cite{neto2009:rev,Armitage:2017cjs,PhysRevLett.102.026802,PhysRevB.75.235423,Armour:2009vj}. While the simplest models of graphene result in massless Dirac fermions, a basic question that has been scrutinized heavily is how and when these low energy excitations can develop a mass gap~\cite{haldane1988:qhe,kane2005:qsh,ryu2009:mass}. Of particular interest in this work is the situation when strong electron-electron interaction drives the mass generation. The most common mechanism is through spontaneous symmetry breaking in which the order parameter couples to a mass term in the Dirac equation. Examples include the formation of N\'eel order \cite{sorella1992:hch,meng2010:qsl} or a  valence bond solid~\cite{li2017:fiqcp,sato2017:dqcp}. Typically 
the critical point between the massless Dirac phase and the symmetry broken state is described by some Gross-Neveu-Yukawa (GNY) field theory~\cite{herbut2006:graphene}, see however~\cite{Ayyar:2014eua,Catterall:2015zua,slagle2015:bil}.  

The simplest route to the two-dimensional massless Dirac equation on a lattice, is through the hopping of electrons,
\begin{equation}\label{eq:hop}
  H_0 = - \sum_{\<i, j\> \alpha} t_{ij} (c^\dagger_{i\alpha} c_{j\alpha} + c^\dagger_{j\alpha} c_{i\alpha}),
\end{equation}
with an appropriately chosen $t_{ij}$, where $c_{i\alpha}$ destroys an electron on lattice site $i$ with spin $\alpha=\uparrow,\downarrow$. In this work we study the hopping matrix elements $t_{ij}$ with a two-dimensional bipartite structure that preserves particle-hole symmetry and is independent of the electron spins. At low energies the electronic structure of such a model is described by spin degenerate ($N_f=2$) four-component massless Dirac fermions. The most celebrated example of such a model is nearest neighbor hopping on a honeycomb lattice with $t_{ij}=t$, which is a basic model for the electronic structure of graphene \cite{neto2009:rev}. Another popular example is a square lattice with nearest neighbor hoppings $t_{ij}=t\eta_{ij}$ where the phases $\eta_{ij}$ realize a $\pi$-flux on each fundamental plaquette \cite{Kogut:1974ag}. To obtain a $\pi$-flux on a square lattice we can choose $\eta_{i,i+ e_x} = 1$ and $\eta_{i,i+ e_y} = (-1)^{i_x}$, where $e_x$ and $e_y$ are the unit vectors in the $x$ and $y$ direction, and $i_x$ is the $x$ component of $i$.

In addition to the usual lattice symmetries and time reversal, $H_0$ possesses certain internal symmetries which will play a central role in our work. Most well-known is the $\SU(2)_s$ spin rotational symmetry which is generated by $\vec{\mathcal S}_i = \frac{1}{2} c^\dagger_{i\alpha}\vec \sigma_{\alpha\beta}c_{i\beta}$. The model also has what is by now a well-known ``hidden''  $\SU(2)_c$ charge symmetry~\cite{yang1990:so4}, which is generated by $\vec{\mathcal C}_i = \frac{1}{2} (\zeta_i (c^\dagger_{i\uparrow}c^\dagger_{i\downarrow} + c_{i\downarrow}c_{i\uparrow}), -\i\zeta_i (c^\dagger_{i\uparrow}c^\dagger_{i\downarrow} - c_{i \downarrow}c_{i \uparrow}), c^\dagger_{i\uparrow}c_{i\uparrow} + c^\dagger_{i\downarrow}c_{i\downarrow} - 1)$,  with $\zeta_i = \pm 1$ depending on the A and B sublattices. Finally, 
$H_0$ has an additional 
$\mathbb{Z}^{sc}_2$ spin-charge flip symmetry 
under which $c_{i\downarrow} \mapsto \zeta_i c^\dagger_{i\downarrow}$, $c_{i\uparrow} \mapsto  c^\dagger_{i\uparrow}$, and
the generators of spin and charge rotations are interchanged, $\vec {\mathcal S}_i \leftrightarrow \vec {\mathcal C}_i$. The $\SU(2)_s\times \SU(2)_c\times \mathbb{Z}^{sc}_2$ symmetries can be combined into an  $\O(4)$ symmetry~\cite{SM}. This $\O(4)$ symmetry is most manifest when the hopping Hamiltonian is rewritten in terms of real ``Majorana'' modes $\gamma_i^a$, with $a=1,2,3,4$, and $\gamma_i^a$ transforming in the vector representation of $\O(4)$, the hopping taking the form, $H_0 = \sum_{\<i, j\> a} \frac{\i}{2} t_{ij}  \gamma^a_i \gamma^a_j$.

Most often electron-electron interactions are added to \cref{eq:hop} by the Hubbard-$U$ term, $H_U = U \sum_i (n_{i\uparrow} - \frac{1}{2}) (n_{i\downarrow} - \frac{1}{2})$. This term preserves the $\SU(2)_s$ and $\SU(2)_c$ symmetries but being odd under the $\mathbb{Z}^{sc}_2$ acts like an Ising magnetic field that breaks the spin-charge flip symmetry. It is well known that repulsive-$U$ interactions favor an antiferromagnetic (AFM) ``spin'' order parameter $\vec \phi^s$, and attractive-$U$ favors a combined charge-density wave (CDW)/superconducting ``charge'' order parameter $\vec \phi^c$. We can understand how these orders couple to the fermions in a simple mean-field model, $H_\mathrm{MF} = H_0 + \sum_i \zeta_i ( \vec {\phi}^s_i \cdot \vec{\mathcal S}_i + \vec {\phi}^c_i \cdot \vec{\mathcal C}_i)$. For $U>0$, $\vec \phi^s \neq 0$ but $\vec \phi^c = 0$ (and vice versa for $U<0$). Since $H_0$ realizes a $2+1$ dimensional  Dirac dispersion, long range order sets in at a finite-$|U|$ phase transition which is described by the so-called ``chiral-Heisenberg'' GNY fixed point that has been the subject of intense numerical~\cite{sorella2012:hchubb,PhysRevX.3.031010,PhysRevX.6.011029,PhysRevB.102.235105} and field theoretic studies~\cite{Gracey:1990wi,Rosenstein:1993zf,janssen2014:frg,zerf2017:4loop}. In this work we look into the nature of the quantum critical phenomena when we add electron-electron interactions to \cref{eq:hop} that preserve the full $\O(4)$ symmetry of the hopping problem, including the crucial $\mathbb{Z}^{sc}_2$ symmetry, which is absent in the usual Hubbard formulation.

Clearly the full $\O(4)$ symmetry of \cref{eq:hop} will be preserved if we add interactions that depend only on  $\sum_\alpha (c^\dagger_{i\alpha} c_{j\alpha} + c^\dagger_{j\alpha} c_{i\alpha})$ with $i, j$ on opposite sublattices.  To this end we focus on a sign-problem-free ``designer Hamiltonian'' (in natural units) which satisfies this criterion,
\thinmuskip=0mu
\medmuskip=1mu 
\thickmuskip=2mu 
\begin{align}
  H_\mathrm{SC} &= - 
  \sum_{\<i, j\>} \exp \Big(\kappa \eta_{ij} \sum_{\alpha=\uparrow, \downarrow} (c^\dagger_{i\alpha} c_{j\alpha} + c^\dagger_{j\alpha} c_{i\alpha})\Big).
\label{spin-charge}
\end{align}
\thinmuskip=3mu
\medmuskip=4mu 
\thickmuskip=5mu
Our model may be viewed as an interacting Hubbard-like model (identical Hilbert space) but with  spin-charge flip symmetry present. Note that our model can be written as a sum of terms defined on bonds of the lattice that consist of fermion bilinears and 4-, 6- and 8-fermion interactions (but no higher order terms) \cite{SM}. For $\kappa\ll 1$ the fermion bilinear terms reproduce \cref{eq:hop} with $t_{ij} = \kappa \eta_{ij}$, and since fermion interactions are perturbatively irrelevant at the massless fixed point the semi-metal phase must emerge at  $\kappa \ll 1$. For  $\kappa \gg 1$, we show using numerical simulations that the Dirac fermions acquire a mass, but because of the spin-charge flip symmetry {\em both} $\vec \phi^s$ and  $\vec \phi^c$ are degenerate and the system breaks the $\mathbb{Z}^{sc}_2$ symmetry by picking one of the two ground states. We present numerical evidence below that the phase transition between Dirac semi-metal and spin-charge flip broken phase is continuous and in a new universality in which both order parameters are simultaneously quantum critical.
We note that other models  preserving the spin-charge flip symmetry include a four-fermion model ~\cite{Li:2019acc,SM} and various fermion-boson models~\cite{assaad2016:dp,goetz2021:ssh}, although they do not harbor the new critical point.

Our designer Hamiltonian \cref{spin-charge} was chosen because we can adapt a fermion bag QMC algorithm to study it \cite{Huffman_2017,Huffman_2020}. By renormalization group (RG) arguments, \cref{spin-charge} is expected to capture universal aspects of the new quantum critical point~\cite{kaul2013:qmc}. The fermion bag algorithm is applicable to all Hamiltonians that are made up of only local terms whose fermionic degrees of freedom are exponentiated bilinears. While this is a limited family of systems, the algorithm is very efficient within its scope of applicability \cite{Huffman_2017}. We expand the partition function $Z = \tr \e^{- H_{SC}/T}$  as
$Z = \sum_k \int \left[\d\tau\right] (-1)^k \Tr \left[H_{SC}(\tau_k) \cdots H_{SC}(\tau_2) H_{SC}(\tau_1)\right].$
Here the notation $\int \left[d\tau\right]$ denotes time-ordered integration for times $1/T \geq \tau_k \geq \cdots \geq \tau_2 \geq \tau_1 \geq 0$.  The expansion can be derived from the continuous-time interaction representation where $H_0 = 0$ and $H_{\rm int} = H_{SC}$ \cite{PhysRevLett.81.2514,PhysRevB.72.035122,PhysRevLett.97.076405,Iazzi_2015,Burovski_2008,PhysRevA.82.053621}, and also resembles the stochastic series expansion \cite{PhysRevB.43.5950,Wang_2016}. The algorithm then involves exploring a configuration space made up of the terms in the expansion, and makes use of 
locality to compute transition probabilities as small determinants \cite{Huffman_2017,Huffman_2020}.  With two spin species, it is immediately evident that there is no sign problem in the
expansion, because every term in the sum is the square of a real number. However, we note that even in models of the form \cref{spin-charge} but with an odd number of flavors there is still no sign problem \cite{Huffman_2014,Li:2014tla,doi:10.1146/annurev-conmatphys-033117-054307,wei2018semigroup}. 
We compute two correlation functions of order parameters,
\begin{align}
    C_S  =2\langle {\cal S}^z_{i_0} {\cal S}^z_{i_1} \rangle, \quad C_U &= \langle {\cal U}^z_{i_0} {\cal U}^z_{i_1} \rangle ,
    \label{qmcobs}
\end{align}
where $C_S$ measures the N\'eel order through the anti-ferromagnetic spin order parameter ${\cal S}^z_i$ and $C_U$ measures the breaking of the spin-charge symmetry through the order parameter ${\cal U}_i =(n_{i\uparrow}-\frac{1}{2}) (n_{i\downarrow}-\frac{1}{2})$, which is a four-fermion operator that is odd under $\mathbb{Z}_2^{sc}$, but invariant under $\SU(2)_s \times \SU(2)_c$. 
In \cref{qmcobs} $i_0 = (0, 0)$ and $i_1 = (L/2,0)$ and we assume $L/2$ is even.
We work at a finite inverse temperature $1/T=L$ and for numerical convenience we henceforth work with the tuning parameter $\mathbf{g}=2\tanh\frac{\kappa}{2}$ instead of $\kappa$~\cite{SM}.

\begin{figure}[t]
    \includegraphics[width=0.45\textwidth]{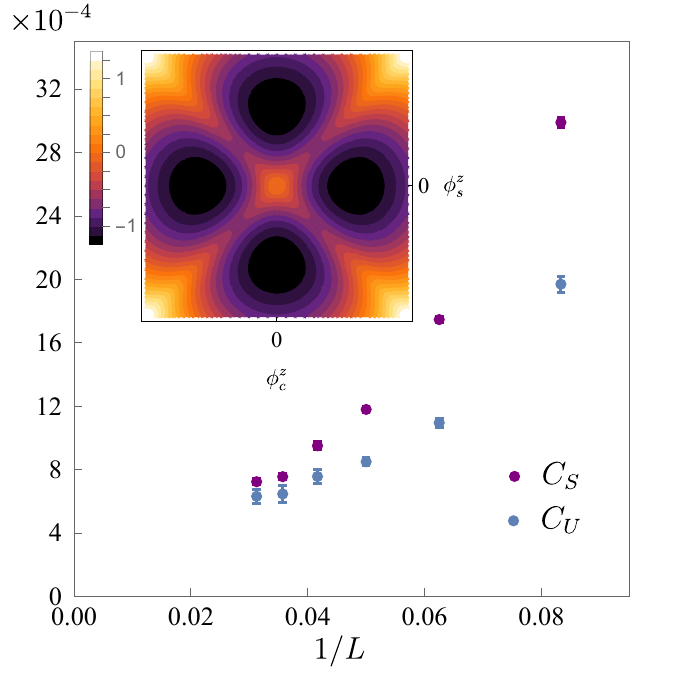}
    \caption{\label{fig:massive} Characterization of the massive phase from QMC and field theory. The main panel shows finite size scaling data for $C_S$ and $C_U$ using the fermion bag QMC method for a coupling constant $\mathbf{g}=1.6$. Both correlation functions scale to a finite value in the thermodynamic limit indicating that the system breaks the $\mathbb{Z}_2^{sc}$ Ising symmetry as well as the SU(2) symmetry of spin and charge. The inset panel shows the effective potential for $\phi_s^z$ and $\phi_c^z$, with $\vec \phi_{s,c}=(0,0,\phi^z_{s,c})$, when these order parameters are coupled to free massless Dirac fermions using the Yukawa coupling, \cref{eq:LY}. Since the minimum of the potential are along the $x$ and $y$ axes, we conclude that the system condenses either $\vec \phi_{s}$ or $\vec{\phi}_{c}$ but not both, consistent with our interpretation of the QMC correlation functions.
    }
\end{figure}

\begin{figure}
    \includegraphics[width=0.45\textwidth]{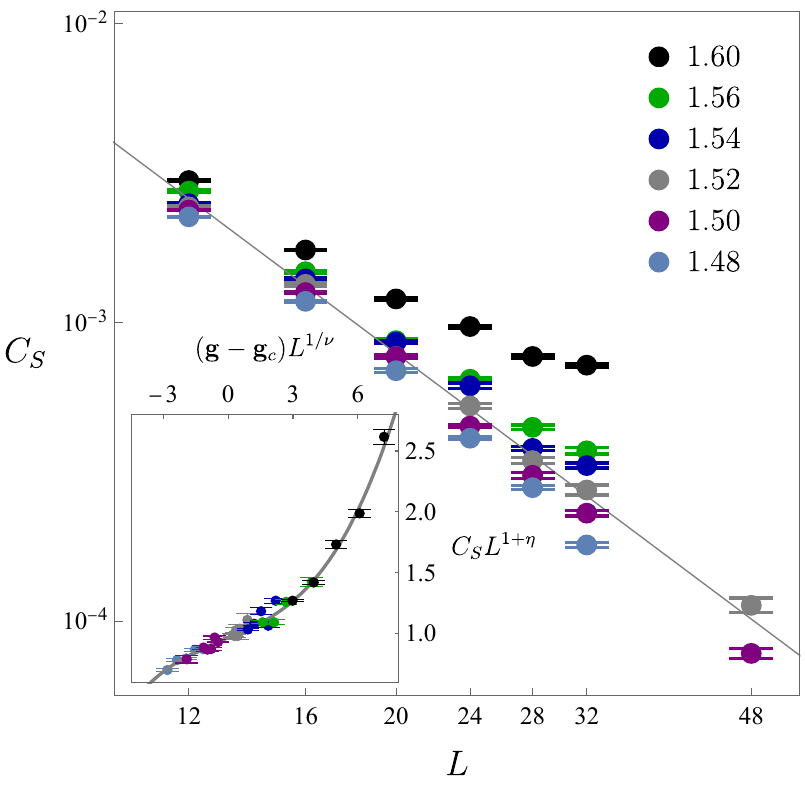}
  \caption{The main plot shows $C_S$ as a function of $L$ on a log-log scale up to $L=48$ for various values of $\mathbf{g}$. For large values of $L$ we find that $C_S$ decays to zero when $\mathbf{g} = 1.48$, while it saturates to a constant when $\mathbf{g}=1.60$, with a phase transition around $\mathbf{g}_c \approx 1.52$ where the data fits to $C_S \approx 0.67/L^{2.25}$ (straight line in the plot). The inset shows that all of the data (after dropping $L=12$) collapse to the universal scaling function discussed in the text with $\eta=1.38(6)$, $\nu=0.78(7)$, and $\mathbf{g}_c=1.514(8)$, providing compelling evidence for a quantum critical point.
  }
   \label{fig:O4}
\end{figure}

We first investigate the nature of the massive phase in our lattice model, $H_{\rm SC}$, using the QMC method described above. We work at a coupling $\mathbf{g} = 1.6$, which is deep in the massive phase. As shown in \cref{fig:massive}, we find a finite value of $C_U$ in the thermodynamic limit, which indicates that the Ising symmetry, $\mathbb{Z}^{sc}_2$ is spontaneously broken. Further we find that $C_S$ also scales to a finite value in the thermodynamic limit, which implies N\'eel order. Together we interpret this to imply that the system has to spontaneously choose between the charge and the spin sector, breaking $\mathbb{Z}^{sc}_2$, and forming either a N\'eel state or a superconductor/CDW state which breaks the corresponding SU(2) symmetry. Next, using QMC we study the nature of the phase transition between the Dirac semi-metal and the massive phase. \cref{fig:O4} shows the data for $C_S$ as a function of system size $L$. For large values of $L$, there is clear evidence that $C_S$ 
converges to a non-zero constant at the coupling $g=1.6$ (massive phase), while it  scales to zero at the coupling $g=1.48$ (Dirac semi-metal). A good fit to the power-law $C_s = 0.67/L^{2.25}$ for $12 \leq L \leq 48$ with a $\chi^2 = 0.95$ is found at the coupling $\mathbf{g}=1.52$ as expected at a quantum critical point. A multi-parameter scaling fit of all our data except for $L=12$, to the form  $C_S = L^{-(1+\eta)} f((g-g_c)\;L^{1/\nu})$ with $f(x) = f_0 + f_1 x + f_2 x^2 + f_3 x^3$ yields $\eta=1.38(6)$, $\nu=0.78(7)$, $\mathbf{g}_c=1.514(8)$, $f_0=0.96(15)$, $f_1=0.073(26)$, $f_2=0.0012(43)$, $f_3 = 0.0026(32)$ with a $\chi^2=1.25$. Interestingly, the large value of $\eta$ clearly establishes that this criticality is not captured by the chiral-Heisenberg theory. We note that $\eta>1$ although uncommon has been observed at certain critical points previously~\cite{isakov2012,janssen2014:frg,boettcher2016}. In the inset of \cref{fig:O4} we show the scaling collapse of this data using the functional form of the multi-parameter fit, providing strong evidence for a continuous quantum critical point.

To capture this observed phenomena, we formulate a field theory in the Euclidean space-time Lagrangian picture~\cite{SM} in terms of the continuum fields 
that are expected to appear as long distance fluctuations near the critical point. 
These are 8-component fermion fields $\bar \psi, \psi$ which are acted upon by tensor products of $4\times 4$ Dirac matrices $\gamma^\mu$ and a spin Pauli matrix $\vec \sigma$. In terms of these fields, the  spin and charge order parameter densities are given by $\vec M_s = \bar \psi_\alpha \vec \sigma_{\alpha\beta} \psi_\beta $ and $\vec M_c = (\psi^{T}_\downarrow\gamma^0 \psi_\uparrow+\bar\psi_\uparrow\gamma^0\bar\psi^{T}_\downarrow ,  i( \psi^{T}_\downarrow\gamma^0 \psi_\uparrow-\bar\psi_\uparrow\gamma^0\bar\psi^{T}_\downarrow ), \bar\psi_\uparrow \psi_\uparrow + \bar\psi_\downarrow \psi_\downarrow) $. Then we can write down the following Yukawa like Lagrangian density,
\begin{align}\label{eq:LY}
  \mathcal{L}_\mathrm{Y} &= -\bar \psi_\alpha \gamma^\mu \partial_\mu \psi_\alpha + g_s \vec \phi_s \cdot \vec M_s + g_c \vec \phi_c \cdot \vec M_c ,
\end{align}
where the first term is the free Dirac theory and the second term describes the interactions of the fermionic fields with critical bosonic fields $\vec \phi_s$ and $\vec \phi_c$ that describe the fluctuations of the anti-ferromagnetic and CDW/superconducting order parameters. In addition to these terms involving the fermionic fields, we supplement our theory with the kinetic terms and self-interactions of the bosonic fields,
\begin{align}
\label{eq:LB}
  {\mathcal L}_{\rm B} & = \sum_{a=s,c}\left ( \frac{1}{2}\partial_\mu \vec\phi_a \cdot \partial^\mu \vec\phi_a + \frac{1}{2} m_a^2 \vec\phi_a \cdot \vec\phi_a + \frac{1}{4!} \lambda_a (\vec\phi_a \cdot \vec\phi_a)^2\right ) \nonumber\\
  &\quad + \frac{1}{12} \lambda_{sc} (\vec\phi_s \cdot \vec\phi_s) (\vec\phi_c \cdot \vec\phi_c).
\end{align}
The first line in the above equation is the usual O(3) $\phi^4$ model for spin and charge sectors. The second line describes a quartic interaction between the spin and charge bosonic fields that is allowed by symmetry. Previous studies of multi-component field theories with fermions have not considered the above model \cite{Roy2011,Roy2014,classen2015:multi,janssen2018:on1n2,Roy2018}.

The Euclidean Lagrangian density ${\mathcal L}_{\rm Y} + {\mathcal L}_{\rm B}$ is expected to describe the critical phenomena in our model.
This theory is symmetric under $\SU(2)_s$ and $\SU(2)_c$; to impose the $\mathbb{Z}^{sc}_2$ we need to require in addition $g_s=g_c$, $\lambda_s=\lambda_c$ and $m_s=m_c$. Then this continuum theory possesses the full $\O(4)$ symmetry of our lattice Hamiltonian, and a thorough analysis of all the Yukawa couplings that are allowed by this $\O(4)$ symmetry can be found in \cite{Liu:2021opm}. When $m_{s,c}$ are large, the bosons will be gapped and we expect a fixed point with massless Dirac particles which we identify with the semi-metal phase in our lattice model. As $m_{s,c}$ is lowered we expect the bosons to condense resulting in a massive phase. Interestingly in this phase the Dirac fermions mediate an interaction between the $\vec \phi_{s,c}$ order parameters. To obtain the effective potential, we assume the condensed bosonic fields are constant in space-time, then we use the O(3) symmetry to rotate $\vec \phi_{s,c}$ fields so they point in the $z$-direction. In this basis, the $\uparrow$ electrons feel a mass $\phi^+ = \phi^z_s + \phi^z_c$ and $\downarrow$ electrons experience  $\phi^- = \phi^z_c - \phi^z_s$. Integrating out the fermions creates an identical effective potential for $\phi^\pm$, which means in the massive phase $\phi^\pm$ condense to the same magnitude but differ at most by a sign (that is determined spontaneously). In the $\phi^z_{c,s}$ language this implies that in the massive phase in the presence of $\mathbb{Z}^{sc}_2$, the  system spontaneously chooses to condense one of $\vec \phi_{s,c}$ and leave the other uncondensed. This is a remarkable mechanism of repulsion between the $\vec \phi_s$ and $\vec \phi_c$ that is generated by the interaction with fermions. The result of an explicit calculation~\cite{SM} of the effective potential from the fermion determinant is plotted in the inset of \cref{fig:massive}, confirming the nature of the massive phase. If $g_s\neq g_c$ the Ising symmetry would be broken and the minima would not be degenerate and the system would then favor the spin (charge) sector as happens in the repulsive (attractive) Hubbard model. 
 \begin{figure}[!t]
  \begin{subfigure}[b]{0.31\textwidth}
    \includegraphics[width=0.97\textwidth]{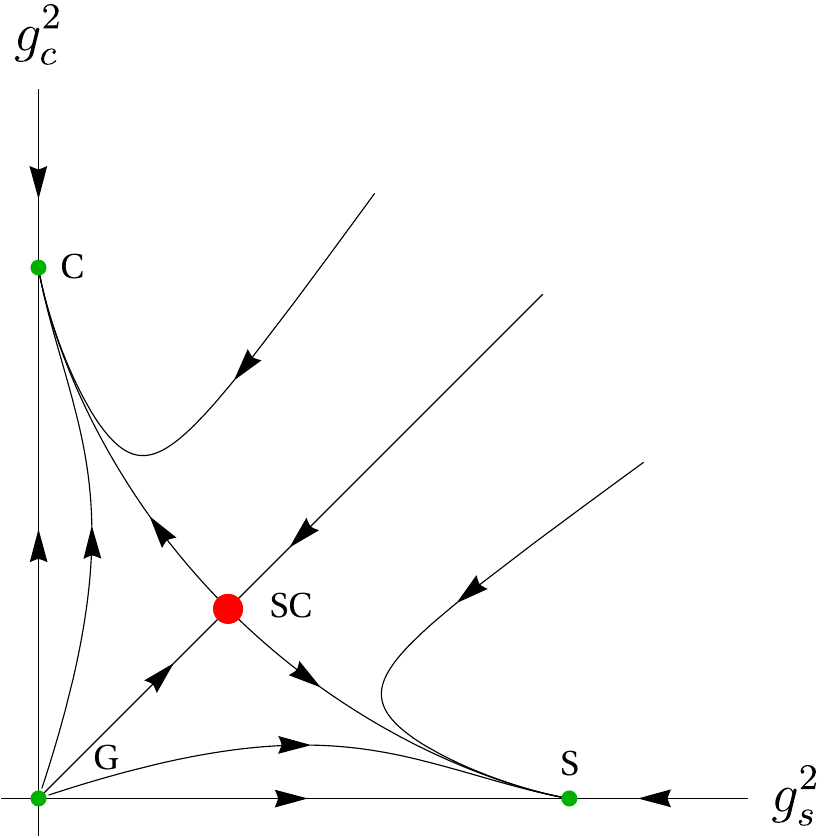}
  \end{subfigure}
  \begin{subfigure}[b]{0.16\textwidth}
    \includegraphics[width=\textwidth]{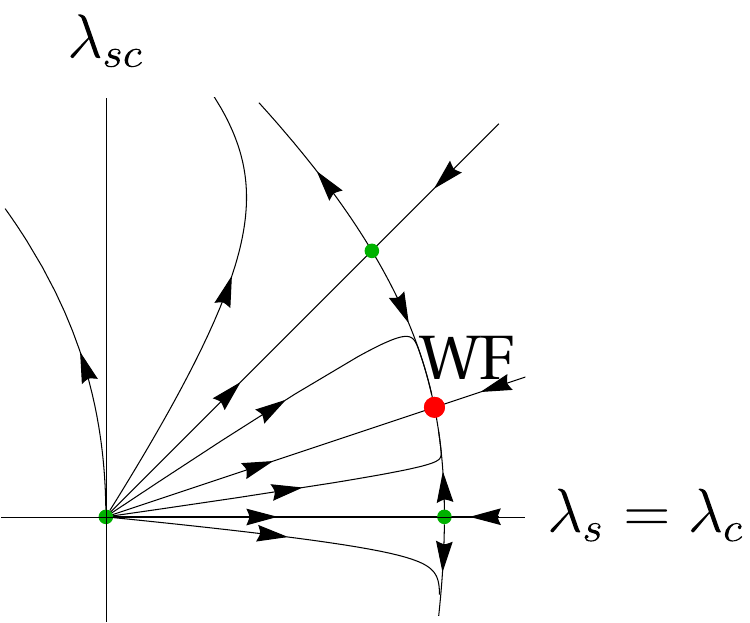}
    \begin{picture}(0,0)
      \put(15,50){$N_f=0$}
    \end{picture}
    \includegraphics[width=\textwidth]{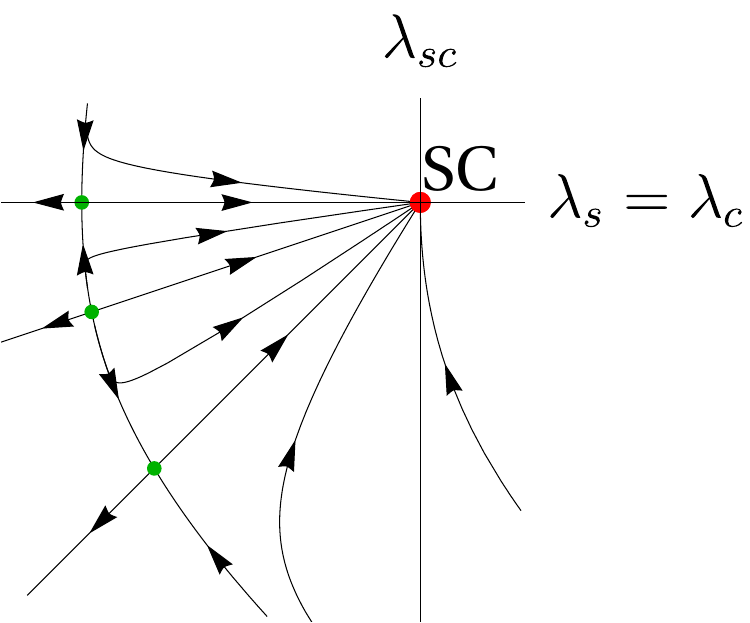}
    \begin{picture}(0,0)
      \put(15,40){$N_f=\infty$}
    \end{picture}
    \label{fig:RG-lambda}
  \end{subfigure}
  \caption{\label{fig:rg}Renormalization group flows of the couplings in the massless ${\cal L}_Y+{\cal L}_B$ theory close to four dimensions. (left) The one-loop flow of the Yukawa coupling $g_s,g_c$ are independent of the $\lambda$ and are shown. In addition to the unstable Gaussian fixed point (G) there are two chiral-Heisenberg fixed points corresponding to the transition from semi-metal to N\'eel (S) and semi-metal to CDW-superconductor (C). In addition there is a new fixed point (SC), which captures the ``chiral spin-charge symmetric" universality, the focus of our study. In models with $\mathbb{Z}^{sc}_2$ symmetry like $H_{\rm SC}$ of Eq.~\ref{spin-charge} the flow is restricted to the diagonal $g_s^2=g_c^2$ line, ensuring the stability of the SC fixed point and the inaccessibility of S and C in such spin-charge symmetric models. As expected the SC fixed point is unstable to breaking of $\mathbb{Z}^{sc}_2$. \red{(right) flow of the boson self interactions in the spin-charge symmetric sector with $g_{s,c}$ fixed at their spin-charge symmetric fixed point values (SC) for $N_f = 0$, and $N_f = \infty$. Due to the duality between small and large $N_f$, the fixed points and flow directions are opposite to each other. In particular, at large $N_f$, there is a stable fixed point (SC) near the origin.} \blue{Together these graphs show that the SC fixed point is stable in the spin-charge symmetric sector and can hence be reached by tuning just one parameter, the mass of the boson. Being the only such fixed point, this qualifies SC as the universal theory of the quantum critical point found in our numerical work.} }
  \label{fig:rg4eps}
\end{figure}

Since all the  non-linear couplings $\lambda_s, \lambda_c,\lambda_{sc},g_s,g_c$ become marginal in four dimensions, we can study the critical region  of ${\mathcal L}_{\rm Y}+{\mathcal L}_{\rm B}$ using the perturbative RG in $4-\varepsilon$ dimensions. We have obtained identical one-loop flow equations using both dimensional regularization with minimal subtraction~\cite{JZJ} and a soft cutoff method~\cite{vojta2000:dwrg} for the massless theory~\cite{SM},
\begin{align}\label{eq:beta-functions}
  \frac{\d g_s^2}{\d\log\ell} &= \varepsilon g_s^2 - \frac{1}{8\pi^2} \big( (2N_f+1)g_s^4 + 9 g_s^2g_c^2 \big), \nonumber \\
  \frac{\d \lambda_s}{\d\log\ell} &= \varepsilon \lambda_s - \frac{1}{8\pi^2} \Big( \frac{11}{6} \lambda_s^2 + \frac{1}{2} \lambda_{sc}^2 + 4N_fg_s^2\lambda_s - 24N_fg_s^4 \Big), \nonumber \\
  \frac{\d \lambda_{sc}}{\d\log\ell} &= \varepsilon\lambda_{sc} - \frac{1}{8\pi^2} \Big( \frac{5}{6} (\lambda_s + \lambda_c) \lambda_{sc} + \frac{2}{3} \lambda_{sc}^2 \nonumber\\
  &\qquad\qquad~ + 2N_f(g_s^2 + g_c^2)\lambda_{sc} - \redeq{72}N_f g_s^2g_c^2 \Big).
\end{align}
The flow equations for $g_c^2$ and $\lambda_c$ can be obtained by exchanging the $s$ and $c$ subscripts using spin-charge flip symmetry. We now study the fixed points of these flow equations. \blue{We note that we can choose $N_f$ in different ways: in $2+1$d our case of interest $N_f=2$, on the other hand our lattice model when extended to $3+1$d would give $N_f=4$. An $\varepsilon$ expansion with either choice can be formulated, and does not affect our main conclusions except for quantitative estimates for the critical exponents. We proceed with $N_f=2$. }Since the $g_{s,c}$ equations do not involve the quartic boson interactions, we can solve them separately. The Yukawa flow equations have four zeros in $(g_s, g_c)$: $(0,0)$ is the unstable Gaussian fixed point (G), $(\frac{4}{\red{N_f+5}}\pi^2\varepsilon, \frac{4}{\red{N_f+5}}\pi^2\varepsilon)$ is a spin-charge symmetric fixed point (SC), $(0, \frac{8}{\red{2N_f+1}}\pi^2\varepsilon)$ (S) and $( \frac{8}{\red{2N_f+1}}\pi^2\varepsilon, 0)$ (C) are chiral-Heisenberg fixed points, as shown in \cref{fig:rg}.  We now ask whether there is a stable spin-charge symmetric fixed point in the bosonic sector when $g_{s,c}$ are evaluated at the SC values. \blue{Indeed with this evaluation there are four bosonic fixed points as shown in \cref{fig:rg}, but only one is stable, providing us with a unique fixed point that captures the universal aspects of the quantum critical point in our lattice model.} 
\red{Based on Eq.~(\ref{eq:beta-functions}), the stable fixed point at $N_f=0$ (Wilson-Fisher) continues to remain stable for small $N_f$ 
and disappears beyond $N_f\approx 0.016$. In addition, we find that there is a stable fixed point in the bosonic sector in the large $N_f$ limit as well, whose existence can be argued using a ``duality" in our one-loop flow equations of the bosonic sector when $g_{s,c}$ is at the SC fixed point and $\lambda_s = \lambda_c$. This duality implies that under the mapping $\lambda_i \mapsto -\lambda_i$ and $N_f \mapsto \frac{25}{N_f}$, the flow diagram at $N_f$ and $\frac{25}{N_f}$ have fixed points and flow directions that are ``flipped" as shown on the right in \cref{fig:rg}. Since in the limit $N_f \rightarrow 0$ fermions decouple and the Gaussian fixed point is fully unstable, in the limit $N_f \rightarrow \infty$ the Gaussian fixed point becomes fully stable. This stable fixed point is unique and moves away from the origin for finite but large $N_f$ and can be found at
$\lambda_s = \lambda_c \sim \frac{48 \pi^2 \varepsilon}{N_f},\quad \lambda_{sc} \sim \frac{144 \pi^2 \varepsilon}{N_f}$. This stable fixed point, that continues to exist until $N_f \approx 16.83$, or the one at small-$N_f$, could provide a qualitative understanding of the quantum critical point that we find using our Monte Carlo method. For $N_f = 2$ of our model, the leading order analysis presented here does not show the existence of a stable fixed point, but it is possible that a higher order analysis in the $\varepsilon$ expansion or large $N_f$ expansion may provide a more quantitatively understanding of the critical point.} By studying the field strength renormalization and the flow of the boson mass term, we can also identify the critical exponents $\eta$ and $\nu$ at large-$N_f$. The critical exponents at one-loop order in $4-\varepsilon$ expansion are~\cite{SM}, $\eta = \redeq{\frac{N_f}{N_f+5}}\varepsilon$, $\eta_\psi = \frac{3}{\redeq{2(N_f+5)}}\varepsilon$, $ \frac{1}{\nu} = 2 - \redeq{\varepsilon}$, where for $\frac{1}{\nu}$ the large $N_f$ limit is taken.

\blue{We note that the quantitative agreement for these exponents between the one-loop $\varepsilon$-expansion  and the numerical data is not great. But such discrepancy has been seen in other GNY type theories and can be attributed in part to the large value of $\eta$.}


RKK acknowledges NSF DMR-2026947, the Aspen Center for Physics (NSF Grant PHY-1607611) and G. Murthy for discussions. SC and HL are supported by the U.S. Department of Energy, Office of Science, Nuclear Physics program under Award No.DE-FG02-05ER41368. Research of EH at the Perimeter Institute is supported in part by the Government of Canada through the Department of Innovation, Science and Economic Development and by the Province of Ontario through the Ministry of Colleges and Universities. \red{The authors thank F. Assaad and L. Janssen for comments on the manuscript, and in particular, I. Herbut and M. Scherer for valuable discussions and pointing out a missing factor of 3 in the flow equation \cite{Herbut:2022zzw} that appeared in the original version of this paper.} This work used computational resources provided by the Extreme Science and Engineering Discovery Environment (XSEDE) \cite{xsede}, which is supported by National Science Foundation grant number ACI-1548562.

\bibliography{Refs}
\clearpage
\onecolumngrid
\appendix

\section{A more conventional form of our lattice Hamiltonian}
In this section we will rewrite the Hamiltonian 
\begin{align}
  H_\mathrm{SC} &= - \sum_{\<i, j\>} \exp \Big(\kappa \eta_{ij} \sum_\alpha (c^\dagger_{i\alpha} c_{j\alpha} + c^\dagger_{j\alpha} c_{i\alpha})\Big)
\tag{\ref{spin-charge}}
\end{align}
in a more conventional form involving polynomials of fermion creation and annihilation operators. We first define and expand
\begin{align}
  H_{\<i, j\>, \alpha} &:= \exp \big(\kappa \eta_{ij} (c^\dagger_{i\alpha} c_{j\alpha} + c^\dagger_{j\alpha} c_{i\alpha})\big) \nonumber\\
  &= \eta_{ij} \sinh \kappa (c^\dagger_{i\alpha} c_{j\alpha} + c^\dagger_{j\alpha} c_{i\alpha}) - 2 (\cosh \kappa - 1) (n_{i\alpha}-\frac{1}{2})(n_{j\alpha}-\frac{1}{2}) + \frac{1}{2}(\cosh \kappa + 1).
\end{align}
Then up to a constant, $H_\mathrm{SC}$ can be written as
\begin{align}
  &\qquad \frac{2}{\sinh \kappa (\cosh \kappa + 1)} H_\mathrm{SC} \nonumber\\
  &= - \frac{2}{\sinh \kappa (\cosh \kappa + 1)} \sum_{\<i, j\>} H_{\<i, j\>, \uparrow} H_{\<i, j\>, \downarrow} \nonumber\\
  &= -\eta_{ij} \sum_\alpha (c^\dagger_{i\alpha} c_{j\alpha} + c^\dagger_{j\alpha} c_{i\alpha}) - \Big(2\tanh\frac{\kappa}{2} \Big) \Big( (c^\dagger_{i\uparrow} c_{j\uparrow} + c^\dagger_{j\uparrow} c_{i\uparrow}) (c^\dagger_{i\downarrow} c_{j\downarrow} + c^\dagger_{j\downarrow} c_{i\downarrow}) - \sum_{\alpha} (n_{i\alpha}-\frac{1}{2})(n_{j\alpha}-\frac{1}{2}) \Big) \nonumber\\
  & \quad + \Big(2\tanh\frac{\kappa}{2} \Big)^2 (\textnormal{6th order}) + \Big(2\tanh\frac{\kappa}{2} \Big)^3 (\textnormal{8th order}) .
\end{align}
Notice that
\begin{align}
  (n_{i\alpha}-\frac{1}{2})(n_{j\alpha}-\frac{1}{2}) = - \frac{1}{2} (c^\dagger_{i\alpha} c_{j\alpha} + c^\dagger_{j\alpha} c_{i\alpha})^2 - \frac{1}{4}.
\end{align}
Hence if we define $2\tanh\frac{\kappa}{2} =: V/t $, then up to a constant we can write
\begin{align}
  \frac{2t}{\sinh \kappa (\cosh \kappa + 1)} H_\mathrm{SC} &= -t \eta_{ij} \sum_\alpha (c^\dagger_{i\alpha} c_{j\alpha} + c^\dagger_{j\alpha} c_{i\alpha}) - \frac{V}{2} \Big( \sum_\alpha (c^\dagger_{i\alpha} c_{j\alpha} + c^\dagger_{j\alpha} c_{i\alpha}) \Big)^2 + (\textnormal{6th and 8th orders}) .
\end{align}
Note that the quadratic and quartic terms of this Hamiltonian is identical with the one in \cite{Li:2019acc}, with $V = J/2$ in their notation. In our work we have defined the coupling $\mathbf{g} = V/t = 2\tanh\frac{\kappa}{2}$ for convenience.

\section{Field theory derivation}

For convenience we approximate the mean field Hamiltonian that describes the order parameter fluctuations of our lattice model through the simplified lattice Hamiltonian,
\begin{align}
\label{eq:MFlatt}
  H_\mathrm{MF} =  - \sum_{\<i, j\> \alpha} t_{ij} (c^\dagger_{i\alpha} c_{j\alpha} + c^\dagger_{j\alpha} c_{i\alpha}) + \sum_i \zeta_i \left( \vec {\phi}^s_i \cdot \vec{\mathcal S}_i +\vec {\bf \phi}^c_i \cdot \vec{\mathcal C}_i \right) .
\end{align}

We take the continuum limit of the above Hamiltonian and use that to derive an effective action in the continuum. Define $\psi_\alpha=(c_{Q_1\alpha},c_{Q_2\alpha},c_{Q_3\alpha},c_{Q_4\alpha})^T$ and $\psi=(\psi_\uparrow,\psi_\downarrow)$ is an 8 component continuum field operator, where $Q_1 = (\pi/2,\pi/2), Q_2 = (-\pi/2,\pi/2), Q_3 = -Q_1, Q_4 = -Q_2$. We arrive at the following Hamiltonian density, 
\begin{align}
\label{eq:MFcont}
  H_\mathrm{MF}^\textnormal{cont} = - \int\d^2x ~& \i \psi_\alpha^{\dagger} (\Gamma^1\partial_1 + \Gamma^2\partial_2) \psi_\alpha + \vec \phi_s \cdot \vec M_s + \vec \phi_c \cdot \vec M_c,
\end{align}
where $\Gamma^1 = \sigma^3\otimes\sigma^3$ and $\Gamma^2 = \sigma^3\otimes\sigma^1$ come from the lattice structure, and 
$\vec M_s = \psi_\alpha^\dagger \gamma^0 \vec \sigma_{\alpha\beta} \psi_\beta $ and $\vec M_c = (\psi^{T}_\downarrow\gamma^0 \psi_\uparrow+\psi_\uparrow^\dagger \gamma^0\psi^{\dagger T}_\downarrow ,  i( \psi^{T}_\downarrow\gamma^0 \psi_\uparrow-\psi_\uparrow^\dagger \gamma^0\psi^{\dagger T}_\downarrow ), \psi_\uparrow^\dagger \gamma^0 \psi_\uparrow + \psi_\downarrow^\dagger \gamma^0 \psi_\downarrow) $
are the same as the main text but written in terms of $\psi^\dagger$ instead of $\bar \psi$ and  $\gamma^0 = \sigma^1 \otimes \mathbbm{1}$. We can see that the continuum limit \cref{eq:MFcont} is consistent with our lattice model \cref{eq:MFlatt}, because the four components of the Dirac field are at momenta $Q_{1,2,3,4}$,  it is easy to see that the matrix $\gamma_0$ connects the momenta $Q_1$ and $Q_3$ (and the pair $Q_2$ and $Q_4$). Note that in our construction $Q_1$ and $Q_3$ satisfy two important properties: $Q_1=-Q_3$ and $Q_1=Q_3+(\pi,\pi)$ (same for $Q_2$ and $Q_4$). Now it is clear why the same $\gamma_0$ appears with superconducting, N\'eel and CDW order parameters that appear in \cref{eq:MFlatt}: It must appear for the SC order parameters because they have the form $c^\dagger_{\bf k\uparrow}c^\dagger_{-\bf k\downarrow}$ and thus because $Q_1=-Q_3$ and $Q_2=-Q_4$, $\gamma_0$ must appear to connect them. On the other hand for CDW and N\'eel can both be thought of charge and spin density waves at $(\pi,\pi)$ with terms such as $c^\dagger_{{\bf k}+{(\pi,\pi)}\uparrow}c_{{\bf k}\uparrow}$, $\gamma_0$ appears because $Q_1=Q_3+(\pi,\pi)$ and $Q_2=Q_4+(\pi,\pi)$.

Using Grassmann coherent fermion path integral, we can rewrite this as the following Lagrangian density
\begin{align}\label{eq:LMF}
  \mathcal{L}_\mathrm{MF} &= -\bar \psi_\alpha \gamma^\mu \partial_\mu \psi_\alpha + \vec \phi_s \cdot \vec M_s + \vec \phi_c \cdot \vec M_c ,
\end{align}
where $\psi_\alpha$ is a 4-component Dirac fermion, $\bar\psi_\alpha = \psi_\alpha^\dagger\gamma^0$, $\alpha = \uparrow,\downarrow$, and $\mu = 1,2,3$. $\gamma^{0,1,2,3,5}$ are five $4\times 4$ Hermitian matrices satisfying the Clifford algebra $\{ \gamma^i, \gamma^j \} = 2 \delta^{ij} \mathbbm{1}_4$. $\gamma^{0,3,5}$ can be chosen to be real, and $\gamma^{1,2}$ to be imaginary. One way consistent with $\Gamma^{1,2}$ coming from the lattice structure is
\begin{align}
  \gamma^0 = \sigma^1 \otimes \mathbbm{1}, \quad \gamma^1 = \sigma^2 \otimes \sigma^3, \quad \gamma^2 = \sigma^2 \otimes \sigma^1, \quad \gamma^3 = \sigma^3 \otimes \mathbbm{1}, \quad \gamma^5 = \gamma^0\gamma^1\gamma^2\gamma^3 = \sigma^2\otimes\sigma^2 . 
\end{align}

When we allow the bosonic fields to fluctuate, we have to include quadratic terms of them,
\begin{align}\label{eq:LGN}
  \mathcal{L}_\mathrm{GN} &= -\bar \psi_\alpha \gamma^\mu \partial_\mu \psi_\alpha + \frac{1}{2g_s^2}\vec\phi_s \cdot \vec\phi_s + \frac{1}{2g_c^2}\vec\phi_c \cdot \vec\phi_c + \vec \phi_s \cdot \vec M_s + \vec \phi_c \cdot \vec M_c ,
\end{align}
which is a Gross-Neveu type model if we integrate out the bosons. On the other hand, when the theory is expand near 4 dimension, the bosons becomes dynamical and self-interactions are generated, and therefore we have the Lagrangian $\mathcal{L}_Y + \mathcal{L}_B$ in the main text.

This form of the Lagrangian is further confirmed by the boson effective potential, which we will calculate next.

\section{The effective potential}
In this section, we derive the effective potential of our continuum model in the broken phase, by integrating out the one-loop fermion determinant in the partition function
\begin{align}
  Z = \int D\psi D\bar\psi D\vec\phi_s D\vec\phi_c ~ \e^{-S[\psi, \bar\psi, \vec\phi_s, \vec\phi_c]}.
\end{align}
Since we will treat the $\vec\phi_s$ and $\vec\phi_c$ fields as constant in space time, using the spin and charge symmetry, we can always rotate $\vec\phi_s$ and $\vec\phi_c$ such that $\phi_s^{1,2} \equiv \phi_c^{1,2} \equiv 0$. Therefore it is sufficient to only consider the third components of spin and charge couplings, and for simplicity, we will denote them by $\phi_s$ and $\phi_c$ in this section. Now the action can be written as
\begin{align}
  S[\psi, \bar\psi, \phi_s, \phi_c] &= \int \d^dx -\bar\psi_\alpha \gamma^\mu\partial_\mu \psi_\alpha + \frac{1}{2g_s^2}\phi_s^2 + \frac{1}{2g_c^2}\phi_c^2 + \phi_s(\bar\psi^1 \psi^1 - \bar\psi^2 \psi^2) + \phi_c(\bar\psi^1 \psi^1 + \bar\psi^2 \psi^2).
\end{align}
Integrating out the fermions, we have
\begin{align}
  S_\eff[\phi_s, \phi_c] &= -\log\Det
           \begin{pmatrix}
             -\slashed\partial + \phi_c + \phi_s & \\
             & -\slashed\partial + \phi_c - \phi_s \\
           \end{pmatrix}
  + \int \d^dx~ \frac{1}{2g_s^2}\phi_s^2 + \frac{1}{2g_c^2}\phi_c^2, \nonumber\\
  &= -\log\Det
           \begin{pmatrix}
             i\slashed k + \phi_c + \phi_s & \\
             & i\slashed k + \phi_c - \phi_s \\
           \end{pmatrix}
  + \int \d^dx~ \frac{1}{2g_s^2}\phi_s^2 + \frac{1}{2g_c^2}\phi_c^2,
\end{align}
where the second line is obtained in the momentum eigenstates of the fermions. Since $\{\slashed k, \gamma^5\} = 0$, while $[\phi_{s,c}, \gamma^5] = 0$, we have
\begin{align}
  \Det
  \begin{pmatrix}
    i\slashed k + \phi_c + \phi_s & \\
    & i\slashed k + \phi_c - \phi_s \\
  \end{pmatrix}
  &= \Det \gamma^5
  \begin{pmatrix}
    i\slashed k + \phi_c + \phi_s & \\
    & i\slashed k + \phi_c - \phi_s \\
  \end{pmatrix} \gamma^5 
  = \Det
  \begin{pmatrix}
    -i\slashed k + \phi_c + \phi_s & \\
    & -i\slashed k + \phi_c - \phi_s \\
  \end{pmatrix} .
\end{align}
Therefore
\begin{align}
  \Det
  \begin{pmatrix}
    i\slashed k + \phi_c + \phi_s & \\
    & i\slashed k + \phi_c - \phi_s \\
  \end{pmatrix}
  &= \Det
  \begin{pmatrix}
    k^2 + (\phi_c + \phi_s)^2 & \\
    & k^2 + (\phi_c - \phi_s)^2 \\
  \end{pmatrix}^2 \nonumber\\
  &= \Det(k^4 + 2k^2(\phi_c^2 + \phi_s^2) + (\phi_c^2 - \phi_s^2)^2)^2  \nonumber\\
  &= \Det(k^4) \Det\Big(1 + 2\frac{\phi_c^2 + \phi_s^2}{k^2} + \frac{(\phi_c^2 - \phi_s^2)^2}{k^4}\Big)^2. 
\end{align}
The factor $\Det(k^4)$ can be ignored because it is independent of $\phi_{s,c}$ and therefore just contributes a constant shift to the effective potential. Now using $\log\Det = \Tr\log$, and interpreting $\Tr$ as integration over momentum eigenstates, we have the following effective potential
\begin{align}
  V_\eff[\phi_s, \phi_c] &= -2\int^\Lambda\frac{\d^d k}{(2\pi)^d}\log\Big(1 + 2\frac{\phi_c^2 + \phi_s^2}{k^2} + \frac{(\phi_c^2 - \phi_s^2)^2}{k^4}\Big) + \frac{1}{2g_s^2}\phi_s^2 + \frac{1}{2g_c^2}\phi_c^2 \nonumber\\
  &= -S_d\int_0^\Lambda\d k^2~k^{d-2}\log\Big(1 + 2\frac{\phi_c^2 + \phi_s^2}{k^2} + \frac{(\phi_c^2 - \phi_s^2)^2}{k^4}\Big) + \frac{1}{2g_s^2}\phi_s^2 + \frac{1}{2g_c^2}\phi_c^2,
\end{align}
where $S_d = \frac{2}{(4\pi)^{d/2}\Gamma(d/2)}$ is the loop integral factor.
For $d = 3$, we have
\begin{align}
  \frac{1}{S_d}V_\eff[\phi_s, \phi_c] &= \frac{2\pi}{3}|\phi_c-\phi_s|^3 -\frac{4}{3}(\phi_c-\phi_s)^3\tan^{-1}\frac{\phi_c-\phi_s}{\Lambda} +\frac{2\pi}{3}|\phi_c+\phi_s|^3 -\frac{4}{3}(\phi_c+\phi_s)^3\tan^{-1}\frac{\phi_c+\phi_s}{\Lambda} \nonumber\\
  &\quad - \frac{8}{3}\Lambda(\phi_c^2 + \phi_s^2) - \frac{2}{3}\Lambda^3\log\Big(1 + 2\frac{\phi_c^2 + \phi_s^2}{\Lambda^2} + \frac{(\phi_c^2 - \phi_s^2)^2}{\Lambda^4}\Big) + \frac{1}{2S_dg_s^2}\phi_s^2 + \frac{1}{2S_dg_c^2}\phi_c^2.
\end{align}
We plotted this effective potential in the broken phase in \cref{fig:massive}, where we set $S_dg_s^2 = S_dg_c^2 = 5$.

\begin{figure}[htp]
  \centering
  \includegraphics[width=0.5\textwidth]{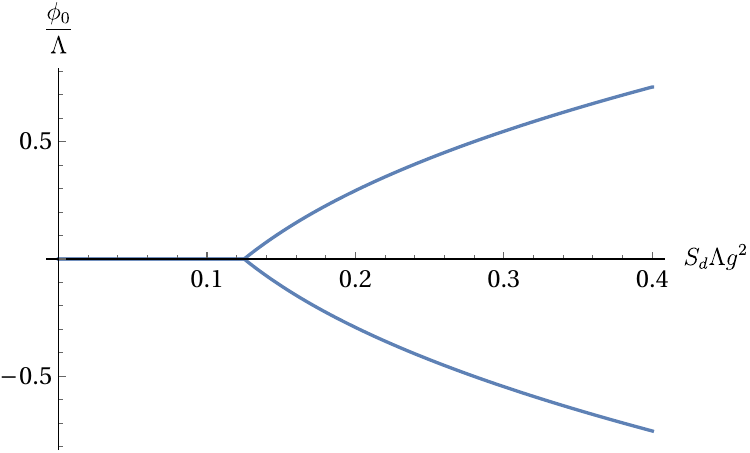}
  \caption{The order parameter $\frac{\phi_0}{\Lambda}$ as a function of $S_d\Lambda g^2$}
  \label{fig:O4-OP}
\end{figure}

We are interested in the minima of the effective potential at $g_s = g_c = g$. Solving the equations,
\begin{align}
  0 &\stackrel{!}{=} \frac{1}{S_d} \frac{\partial V_\eff[\phi_s, \phi_c]}{\partial\phi_s} = 4(\phi_s-\phi_c)^2\tan^{-1}\frac{\Lambda}{\phi_s-\phi_c} +4(\phi_s+\phi_c)^2\tan^{-1}\frac{\Lambda}{\phi_s+\phi_c} + \Big(\frac{1}{S_dg^2} - 8\Lambda \Big)\phi_s \nonumber\\
  0 &\stackrel{!}{=} \frac{1}{S_d} \frac{\partial V_\eff[\phi_s, \phi_c]}{\partial\phi_c} = 4(\phi_c-\phi_s)^2\tan^{-1}\frac{\Lambda}{\phi_c-\phi_s} +4(\phi_s+\phi_c)^2\tan^{-1}\frac{\Lambda}{\phi_s+\phi_c} + \Big(\frac{1}{S_dg^2} - 8\Lambda \Big)\phi_c ,
\end{align}
we see that there is a trivial solution at $\phi_s = \phi_c = 0$, and non-trivial solutions at
\begin{align}
  \frac{\phi_s \pm \phi_c}{\Lambda} \tan^{-1}\frac{\Lambda}{\phi_s \pm \phi_c} = 1 - \frac{1}{8S_d \Lambda g^2} .
\end{align}
Since $x\tan^{-1}\frac{1}{x}$ has value between $0$ and $1$, and is monotonic for $x>0$, we know that there is a non-trivial solution only when $8S_d\Lambda g^2 \geq 1$, and therefore we can define $g_\crit^2 = \frac{1}{8S_d\Lambda}$.
Let's denote this solution by $\phi_s \pm \phi_c = \phi_0$. Assuming $\phi_s \geq 0$ and $\phi_c \geq 0$, we see that there are three nontrivial solutions: $(\phi_s, \phi_c) = (\phi_0, 0)$, $(\phi_s, \phi_c) = (0, \phi_0)$, $(\phi_s, \phi_c) = (\frac{\phi_0}{2}, \frac{\phi_0}{2})$. The last solution $(\phi_s, \phi_c) = (\frac{\phi_0}{2}, \frac{\phi_0}{2})$ is not a minimum, because the trace of the Hessian
\begin{align}
  \frac{1}{S_d} \Big(\frac{\partial^2 V_\eff[\phi_s, \phi_c]}{\partial\phi_s \partial\phi_s} + \frac{\partial^2 V_\eff[\phi_s, \phi_c]}{\partial\phi_c \partial\phi_c} \Big) = -\frac{8\Lambda}{1+(\frac{\Lambda}{2\phi_s})^2} < 0.
\end{align}
In \cref{fig:O4-OP} we plot the order parameter $\frac{\phi_0}{\Lambda}$ as a function of $S_d\Lambda g^2$.

Finally, when $g \rightarrow g_{\crit+}$, $\phi$ is small, and we can approximate by $\tan \frac{1}{x} \approx \frac{\pi}{2} \sgn(x)$. Then we have 
\begin{align}
  \frac{\pi}{2} \frac{|\phi|}{\Lambda} \approx 1 - \frac{1}{8S_d\Lambda g^2} \implies |\phi_0| \approx \frac{1}{4S_d\pi}\Big( \frac{1}{g_0^2} - \frac{1}{g^2} \Big) = \frac{1}{4S_d\pi} \frac{g^2 - g_0^2}{g_0^4},
\end{align}
from which we can read the critical exponent $\frac{1}{2}(1+\eta)\nu =: \beta = 1$, which agrees with our leading order result from large $N$.

\section{\texorpdfstring{$4-\varepsilon$}{4-epsilon} expansion}
In 4 dimension, the kinetic terms of the scalar fields become marginal, and we have dynamic scalar fields. Therefore we have the Lagrangian $\mathcal{L}_Y + \mathcal{L}_B$. In this section the fields and couplings with subscript $0$ are the bare ones, while those without subscript $0$ are the renormalized ones.  In $4-\varepsilon$ dimension, the bare fields and couplings have dimension $[\psi_0] = \frac{3-\varepsilon}{2}$, $[\phi_0] = \frac{2-\varepsilon}{2}$, $[\lambda_0] = \varepsilon$ and $[g_0] = \frac{\varepsilon}{2}$, and the Lagrangian can be written as
\begin{align}
  \mathcal{L} &= - Z_\psi \bar \psi_\alpha \gamma^\mu \partial_\mu \psi_\alpha + g_{s0} Z_s^{1/2} Z_\psi \vec \phi_s \cdot \vec M_s + g_{c0} Z_c^{1/2} Z_\psi \vec \phi_c \cdot \vec M_c \nonumber\\
  &\quad + \sum_{a=s,c}\left ( \frac{1}{2} Z_a \partial_\mu \vec\phi_a \cdot \partial^\mu \vec\phi_a + \frac{1}{2} Z_a m_{a0}^2 \vec\phi_a \cdot \vec\phi_a + \frac{1}{4!} Z_a^2 \lambda_{a0} (\vec\phi_a \cdot \vec\phi_a)^2\right )  + \frac{1}{12} Z_s Z_c \lambda_{sc0} (\vec\phi_s \cdot \vec\phi_s) (\vec\phi_c \cdot \vec\phi_c) \nonumber\\
  &= - \bar \psi_\alpha \gamma^\mu \partial_\mu \psi_\alpha + m^{\varepsilon/2}g_s \vec \phi_s \cdot \vec M_s + m^{\varepsilon/2}g_c \vec \phi_c \cdot \vec M_c \nonumber\\
  &\quad + \sum_{a=s,c}\left ( \frac{1}{2} \partial_\mu \vec\phi_a \cdot \partial^\mu \vec\phi_a + \frac{1}{2} m_a^2 \vec\phi_a \cdot \vec\phi_a + \frac{1}{4!} m^\varepsilon \lambda_a (\vec\phi_a \cdot \vec\phi_a)^2\right )  + \frac{1}{12} m^\varepsilon \lambda_{sc} (\vec\phi_s \cdot \vec\phi_s) (\vec\phi_c \cdot \vec\phi_c) \nonumber\\
  &\quad - \delta_\psi \bar \psi_\alpha \gamma^\mu \partial_\mu \psi_\alpha + m^{\varepsilon/2}\delta_{g_s} \vec \phi_s \cdot \vec M_s + m^{\varepsilon/2}\delta_{g_c} \vec \phi_c \cdot \vec M_c \nonumber\\
  &\quad + \sum_{a=s,c}\left ( \frac{1}{2} \delta_a \partial_\mu \vec\phi_a \cdot \partial^\mu \vec\phi_a + \frac{1}{2} \delta_{m_a} \vec\phi_a \cdot \vec\phi_a + \frac{1}{4!} m^\varepsilon \delta_{\lambda_a} (\vec\phi_a \cdot \vec\phi_a)^2\right )  + \frac{1}{12} m^\varepsilon \delta_{\lambda_{sc}} (\vec\phi_s \cdot \vec\phi_s) (\vec\phi_c \cdot \vec\phi_c),
\end{align}
from which we can read
\begin{align}\label{eq:counter-terms}
  &\quad Z_\psi = 1 + \delta_\psi, \quad Z_a = 1 + \delta_a, \quad Z_am_{a0}^2 = m_a^2 + \delta_{m_a}, \nonumber\\
  g_{a0} Z_a^{1/2} Z_\psi &= m^{\varepsilon/2} (g_a + \delta_{g_a}), \quad Z_a^2 \lambda_{a0} = m^\varepsilon (\lambda_a + \delta_{\lambda_a}), \quad Z_sZ_c \lambda_{sc0} = m^\varepsilon (\lambda_{sc} + \delta_{\lambda_{sc}}), 
\end{align}
where $a = s, c$, and $\alpha = 1, \cdots, N_f$. The counter terms $\delta_i$ are formal series in the renormalized couplings. In the minimal subtraction scheme, $\delta_i$ are determined order by order by canceling the divergent part of the renormalized part at one loop higher. In \cref{fig:mf-renormalization} we show the divergent diagrams which contributes to the mass and field strength renormalization, and in \cref{fig:i-renormalization} we show the divergent diagrams which contributes to the renormalization of the couplings $g_a$, $\lambda_a$ and $\lambda_{sc}$. The dashed lines denote the boson fields, while the solid lines denote the fermion fields. We pack all the diagrams with the same topology together for compactness. For example, in \cref{fig:delta_m}, when the external lines are $\phi_s$, the loop can be either $\phi_s$ or $\phi_c$. Furthermore, due to the charge coupling containing $\psi\psi$ and $\bar\psi\bar\psi$ terms, the fermion number is not conserved, and the possibility of flipping the directions of the fermion propagators needs to be additionally considered for the diagrams shown below.

\begin{figure}[htp]
  \centering
  \begin{subfigure}[b]{0.3\textwidth}
    \feynmandiagram [layered layout, horizontal=a to b] {
      a -- [scalar] b -- [scalar, out=135, in=45, loop, min distance=3cm] b -- [scalar] c ,
    };
    \caption{$\delta_{m_a} = \frac{1}{6\varepsilon} S_d (5\lambda_a + 3 \lambda_{sc}) m_a^2$}
    \label{fig:delta_m}
  \end{subfigure}
  \begin{subfigure}[b]{0.3\textwidth}
    \feynmandiagram [layered layout, horizontal=b to c] {
      a -- [scalar] b
      -- [fermion, half right, looseness=1.5] c
      -- [fermion, half right, looseness=1.5] b,
      c -- [scalar] d,
    };
    \caption{$\delta_{\phi_a} = -\frac{2}{\varepsilon}S_d N_fg_a^2$}
  \end{subfigure}
  \begin{subfigure}[b]{0.3\textwidth}
    \feynmandiagram [layered layout, horizontal=b to c] {
      a -- [fermion] b
      -- [fermion] c
      -- [scalar, half right, looseness=2] b,
      c -- [fermion] d,
    };
    \caption{$\delta_\psi = -\frac{3}{2 \varepsilon} S_d (g_s^2 +g_c^2)$}
  \end{subfigure}
  \caption{Mass and field strength renormalizations}
  \label{fig:mf-renormalization}
\end{figure}

\begin{figure}[htb]
  \centering
  \begin{subfigure}[b]{0.3\textwidth}
    \feynmandiagram [horizontal=a to b] {
      i1 -- [scalar] a -- [scalar] i2,
      a -- [scalar, half left, looseness=1.5] b
      -- [scalar, half left, looseness=1.5] a,
      f1 -- [scalar] b -- [scalar] f2,
    };
    \caption{$\delta_{\lambda_a}^b = \frac{1}{2\varepsilon} S_d ( 3 \lambda_a^2 + \lambda_{sc}^2 )$ \newline \hphantom{(a)} $\delta_{\lambda_{sc}}^b = \frac{1}{6\varepsilon} S_d ( 5 (\lambda_s + \lambda_c) \lambda_{sc} + 4\lambda_{sc}^2 )$}
  \end{subfigure}
  \begin{subfigure}[b]{0.3\textwidth}
    \feynmandiagram [small, horizontal=a to b] {
      i1 -- [scalar] a,
      i2 -- [scalar] b,
      a -- [fermion] b -- [fermion] c -- [fermion] d -- [fermion] a ,
      f1 -- [scalar] c,
      f2 -- [scalar] d,
    };
    \caption{\centering$\delta_{\lambda_a}^f = \frac{24}{\varepsilon} S_d N_f g_a^4$ \newline  $\delta_{\lambda_{sc}}^f = \frac{\red{72}}{\varepsilon} S_d N_f g_s^2 g_c^2$}
  \end{subfigure}
  \begin{subfigure}[b]{0.3\textwidth}
    \feynmandiagram [small, horizontal=c to d] {
      i1 -- [anti fermion] a
      -- [scalar] b
      -- [fermion] c
      -- [fermion] a,
      b -- [anti fermion] i2,
      c -- [scalar] d,
    };
    \caption{\centering $\delta_{g_s} = \frac{1} {\varepsilon} S_d (-g_s^3 + 3g_sg_c^2)$ \newline 
    $ \delta_{g_c} = \frac{1} {\varepsilon} S_d (-g_c^3 + 3g_cg_s^2)$}
  \end{subfigure}
  \caption{Interaction renormalizations. $\delta_\lambda = \delta_\lambda^b + \delta_\lambda^f$, where the superscript $b$ and $f$ means the contribution from the boson loop and the fermion loop respectively.}
  \label{fig:i-renormalization}
\end{figure}

Using \cref{eq:counter-terms}, we can write the bare couplings in terms of the counter terms $\delta_i$. Requiring the bare couplings to be RG invariant, we get a set of linear equations of the $\beta$ functions, which gives
\begin{align}
  \frac{\d g_s^2}{\d\log\ell} &= \varepsilon g_s^2 - S_d \big( (2N_f+1)g_s^4 + 9 g_s^2g_c^2 \big), \nonumber\\
  \frac{\d \lambda_s}{\d\log\ell} &= \varepsilon \lambda_s - S_d \Big( \frac{11}{6} \lambda_s^2 + \frac{1}{2} \lambda_{sc}^2 + 4N_fg_s^2\lambda_s - 24N_fg_s^4 \Big), \nonumber\\
  \frac{\d \lambda_{sc}}{\d\log\ell} &= \varepsilon\lambda_{sc} - S_d \Big( \frac{5}{6} (\lambda_s + \lambda_c) \lambda_{sc} + \frac{2}{3} \lambda_{sc}^2 + 2N_f(g_s^2 + g_c^2)\lambda_{sc} - \red{72}N_f g_s^2g_c^2 \Big).\tag{\ref{eq:beta-functions}}
\end{align}
The $\beta$ function of $g_c^2$ and $\lambda_c$ can be obtained by $s \leftrightarrow c$.

The $\beta$ functions for $(g_s, g_c)$ have four zeros: $(0,0)$ is the Gaussian unstable fixed point, $(\frac{\varepsilon}{2S_d (N_f+5)}, \frac{\varepsilon}{2S_d (N_f+5)})$ is a spin-charge symmetric saddle point, $(0, \frac{\varepsilon}{S_d (2N_f+1)})$ and $( \frac{\varepsilon}{S_d (2N_f+1)}, 0)$ are stable fixed points. At the spin-charge symmetric saddle point, i.e. $(g_s^2, g_c^2) = (\frac{\varepsilon}{2S_d (N_f+5)}, \frac{\varepsilon}{2S_d (N_f+5)})$, \blue{the $\beta$ functions for $(\lambda_s, \lambda_c, \lambda_{sc})$ have six zeros, 
see \cref{fig:rg4eps} for the RG flow in the $\lambda_{sc} - (\lambda_s=\lambda_c)$ plane.}
\red{the stable fixed point in $(\lambda_s = \lambda_c, \lambda_{sc})$ in the large $N_f$ limit is 
\begin{align}
  \lambda_s = \lambda_c \sim \frac{6\varepsilon}{S_d N_f}, \quad \lambda_{sc} \sim \frac{18\varepsilon}{S_d N_f} ,
\end{align}
}

The critical exponents are
\begin{align}
  \eta_{\phi_{s,c}} &= 2S_d N_fg_{s,c}^2 , \quad \eta_\psi = \frac{3}{4} S_d (g_s^2+g_c^2), \quad  \nu^{-1} = \eta_m + 2 = 2 - S_d \Big( 2N_f g_{s,c}^2 + \frac{5}{6}\lambda_{s,c} + \frac{1}{2} \lambda_{sc} \Big).
\end{align}
At the spin-charge symmetric fixed point \blue{$(g_s^2, g_c^2) = (\frac{\varepsilon}{2S_d (N_f+5)}, \frac{\varepsilon}{2S_d (N_f+5)})$ and $\lambda_s = \lambda_c = \lambda_{sc} = \frac{3( 5-N_f + \sqrt{N_f^2 + 46N_f + 25} )}{14S_d (N_f+5)} \varepsilon$}, we have
\begin{align}
  \eta_{\phi_{s,c}} &= \frac{N_f}{N_f+5}\varepsilon, \quad \eta_\psi = \frac{3}{2(N_f+5)}\varepsilon, \quad \nu^{-1} \sim 2 - \red{\frac{N_f+14}{N_f}\varepsilon \sim 2 - \varepsilon}.
\end{align}
\blue{For $N_f = 2$ and $\varepsilon = 1$, we have
\begin{align}
  \eta_{\phi_{s,c}} &= 0.29 , \quad \eta_\psi = 0.21, \quad \nu = 0.88.
\end{align}
}

\section{QMC Data}
Our QMC calculations have been performed in continuous time using the fermion bag approach as referred to in the main paper. The correlation function data for the observables $C_S$ and $C_U$ are given by Table \ref{datatable2}. This data was used in Figs 1, 2 and the critical fits explained in the text.

\begin{table}[htb]
\begin{center}
\small
\begin{tabular}{|l||l|l|l|l|l|l|l|}
\hline
 $\mathbf{g}$ & $L=12$ & $L=16$ & $L=20$ & $L=24$ & $L=28$ & $L=32$ & $L=48$ \\
 \hline
\multicolumn{8}{|c|}{$C_S$} \\
\hline
$1.48$ & 0.00226(2) & 0.00118(2) & 0.00069(2) &
0.00041(1) & 0.000279(8) & 0.000182(7) & $\qquad -$\\
\hline
$1.50$ & 0.00238(3) & 0.00126(2) & 0.00077(2) &
0.00045(1) & 0.00031(1) & 0.00023(1) &  0.000078(6)\\
\hline
$1.52$ & 0.00244(4) & 0.00134(4) & 0.00078(2) & 0.00053(2) & 0.00035(2) & 0.00027(2) & 0.00011(1)\\
\hline
$1.54$ & 0.00250(4) & 0.00140(3) & 0.00086(2) & 0.00061(3) & 0.00038(1) & 0.00033(1) & $\qquad -$ \\
\hline
$1.56$ & 0.00276(6) & 0.00148(3) & 0.00087(3) & 0.00065(2) & 0.00045(2) & 0.00037(2) & $\qquad -$ \\
\hline
$1.60$ & 0.00299(3) & 0.00172(2) & 0.00113(3) & 0.00089(4) & 0.00071(3) & 0.00068(3) & $\qquad -$ \\
\hline
\multicolumn{8}{|c|}{$C_U$} \\
\hline
$1.60$ & 0.00197(5) & 0.00110(3) & 0.00085(3) & 0.00076(5) & 0.00065(5) & 0.00063(4) & $\qquad -$ \\
\hline
\end{tabular}
\vspace{.5cm}
\caption{Results for the correlation functions $ C_S$ and $C_U$ defined in (\ref{qmcobs}), for our model near the quantum critical point. A seven parameter fit of the data (omitting $L=12$) gives us $\eta=1.38(6)$, $\nu=0.78(7)$, $\mathbf{g}_c= 1.514(8)$, $f_0=1.0(2)$, $f_1=0.07(3)$, $f_2=0.001(4)$, $f_3=0.003(3)$, with a $\chi^2=1.253$. If we fix the critical coupling to $\mathbf{g}_c=1.52$, the parameter fit gives us $\eta=1.34(2)$, $\nu=0.78(7)$, $f_0=0.87(6)$, $f_1=0.06(2)$, $f_2=0.004(2)$, $f_3=0.002(3)$, with a $\chi^2=1.226$.}
\label{datatable2}
\end{center}
\end{table}

\end{document}